# Observation of Resonance of Kagome Flat Band Doublet

Renjie Zhang[1*], Bei Jiang[1*], Xiangqi Liu[2*], Hengxin Tan[3], Xuefeng Zhang[2], Mojun Pan[1], Quanxin Hu[1], Yiwei Cheng[4,1], Chengnuo Meng[4], Yudong Hu[1], Yufan Zhao[4], Runze Wang[1], Dupeng Zhang[1], Junqin Li[4], Zhengtai Liu[4], Mao Ye[4], Ziqiang Wang[5], Yaobo Huang[4†], Gang Li[2], Yanfeng Guo[2†], Hong Ding[1,6,7†], Baiqing Lv[1,3,8†]

[1] *State Key Laboratory of Micro-nano Engineering Science, Tsung-Dao Lee Institute, Shanghai Jiao Tong University, Shanghai 201210, China*

[2] *State Key Laboratory of Quantum Functional Materials, School of Physical Science and Technology, ShanghaiTech University, Shanghai 201210, China*

[3] *School of Physics and Astronomy, Shanghai Jiao Tong University, Shanghai 200240, China*

[4] *Shanghai Synchrotron Radiation Facility, Shanghai Advanced Research Institute, Chinese Academy of Sciences, 201204 Shanghai, China*

[5] *Department of Physics, Boston College, Chestnut Hill, Massachusetts 02467, USA*

[6] *Hefei National Laboratory, Hefei 230088, China*

[7] *New Cornerstone Science Laboratory, Tsung-Dao Lee Institute, Shanghai Jiao Tong University, Shanghai 200240, China*

[8] *Zhangjiang Institute for Advanced Study, Shanghai Jiao Tong University, Shanghai 200240, China*

\* These authors contributed equally to this work

†huangyaobo@sari.ac.cn

†guoyf@shanghaitech.edu.cn

†dingh@sjtu.edu.cn

†baiqing@sjtu.edu.cn

## Abstract

The interplay between local and itinerant electrons underpins many correlated and topological quantum states. Kagome lattices provide an ideal platform by hosting both flat (localized states) and dispersive bands (itinerant states), yet direct spectroscopic evidence of their dynamical coupling has remained elusive. Here we report the long-sought flat band resonance in the quasi-two-dimensional kagome bilayer material $CsCr_6Sb_6$. Using angle-resolved photoemission spectroscopy, transport measurements, and combined density functional theory and dynamical mean-field theory, we identify coexisting flat band doublets and dispersive bands near the Fermi energy. Upon cooling, the flat and dispersive bands exhibit a pronounced enhancement of spectral weight and hybridization, directly evidencing flat band resonance. Crucially, this emergence coincides with the onset of short-range antiferromagnetic correlations, contrasting sharply with conventional Kondo lattice behavior. Our findings demonstrate not only the long-sought flat band resonance in kagome materials, but also its unconventional correlation with magnetism.

## Introduction

The engineering of quantum correlated and topological phases of matter represents a prominent frontier in condensed matter physics. The interplay between itinerant and local states near the Fermi energy ($E_F$)[1] is known to drive a variety of emergent quantum phenomena, such as quantum critical point[2–5], unconventional superconductivity[4,6], and fractional Chern insulatos[7,8]. A hallmark of the interaction between localized (flat band) and itinerant (dispersive band) states in quantum materials is the so-called flat band resonance, characterized by a significant enhancement in the near-$E_F$ single-particle spectral weight[3].

Practical systems supporting the coexistence of local and itinerant electrons can be categorized into three types, as illustrated in Fig 1(a). Type I materials feature atomic flat bands, primarily formed by $4f/5f$ local electrons. A well-known example is the heavy fermion system, including Ce-/Yb-/U-based $f$-electron compounds[9]. This system serves as a paradigm for studying unconventional superconductivity and quantum critical points, understood as arising from the competition between the Ruderman–Kittel–Kasuya–Yosida (RKKY) interaction and the Kondo interaction[3,4,9]. Type II materials consist of van der Waals heterostructures, where flat band resonance emerges either from moiré interference or from the direct combination of local and itinerant components[10–12]. Although these heterostructures offer greater tunability, their investigation is constrained by the complexity of fabrication. Type III materials exploit quantum geometric frustration within specific lattices to generate flat bands[13,14]. Representative examples include the kagome[15–17,7,8,18], Lieb[19], and dice lattices[20]. Compared to other categories, type III materials host two distinct advantages simultaneously: broad accessibility and high tunability. First, due to the universality of flat band engineering via geometry frustration, a wider range of atomic orbitals (e.g., *s*- or *d*-orbitals) can contribute to flat band formation. Remarkably, theoretical predictions[21,22] have identified thousands of candidate materials exhibiting coexisting frustrated flat and dispersive bands, significantly expanding the possibilities for realizing flat band resonance and tuning the competition between RKKY and Kondo interactions. Additionally, the absence of complex fabrication requirements makes type III materials more amenable to spectroscopic investigations, enabling detailed studies of flat band phenomena. Second, akin to type II materials, many type III materials exhibit high tunability, allowing their electronic structure to be controlled via methods, including gating, mechanical exfoliation, and strain—essential tools for exploring the intricate interplay between flat and dispersive bands. Experimentally, angle-resolved photoemission spectroscopy (ARPES) — an experimental technique based on the photoelectric effect — is arguably the most powerful method for probing the flat and dispersive bands as well as their interplay. Despite the great efforts, the spectroscopic evidence of flat band resonance is lacking.

Among the proposed geometric frustration system, the kagome lattice has recently attracted extensive attention as a more accessible and versatile platform for exploring flat band resonance and novel correlated quantum phases[4,5,23–25] due to the coexistence of flat and dispersive bands[15,16,13,26–38]. To realize the desired flat band resonance in kagome materials, we believe several criteria must be met: 1) the presence of narrow flat bands near $E_F$, where Coulomb interactions dominate over kinetic energy, thereby dictating the electronic structure; 2) the coexistence of dispersive bands, which facilitate interactions between flat and itinerant bands; 3) weak interlayer coupling and ideal 2D characteristics of the flat bands, leading to a divergent density of states and therefore the potential for novel quantum states; and 4）the existence of stable local magnetic moments in the absence of long-range magnetic order.

Recently, a new kagome material, $CsCr_6Sb_6$, has been synthesized[39,40], and it meets all the necessary criteria for observing flat band resonance. First of all, $CsCr_6Sb_6$ consists of kagome bilayer units and exhibits limited interlayer coupling due to relatively large interlayer distances, making it an ideal candidate for hosting 2D flat bands and the intrinsic intersection between the kagome flat bands and dispersive bands. Second, it displays a large Sommerfeld coefficient, indicating the existence of near-$E_F$ flat band with large effective mass. Last but not least, the material exhibits the coexistence of short-range antiferromagnetism (AFM) and a Kondo-like logarithmic divergence in resistivity (Fig. 1b). These characteristics point towards the existence of kagome flat band resonance in $CsCr_6Sb_6$.

In this work, taking the advantage of the high-resolution ARPES, we directly observe the coexistence and intersection of flat and dispersive bands near the $E_F$, originating from the kagome doublet bands of the bilayer unit. By performing density functional theory (DFT) and DFT+ dynamical mean field theory (DMFT) calculations, we reveal that the strongly correlated electronic structure of $CsCr_6Sb_6$ near $E_F$ is predominantly derived from Cr $d$-orbitals, in sharp contrast to the extensively studied 135-type of kagome materials, such as $AT_3Sb_5$ ($A$ = alkaline metal, $T$ = V, Cr). More importantly, we uncover an incoherence-to-coherence crossover of the flat bands that coincides with the emergence of weak short-range AFM order around 72 K. This flat band resonance deviates from the traditional Kondo lattice model, where the Kondo resonance typically precedes and competes with the long-range AFM order. Instead, our findings suggest that AFM fluctuations intrinsic to the kagome lattice suppress long-range order while simultaneously enabling flat band resonance. These fluctuations also disrupt Kondo singlets, leading to a synchronized emergence of resonance and short-range AFM. Our results not only provide compelling evidence for the long-sought flat band resonance in kagome materials but also underscore the kagome lattice as a promising platform for studying the interplay between flat band, dispersive bands and enhanced magnetic fluctuations.

**Results**

**Coexistence of Flat Bands and Dispersive Bands**

Like other members of the kagome 166 family, such as $CsV_6Sb_6$ [41], $CsCr_6Sb_6$ crystallizes in a van der Waals-like layered structure with the space group $R$-$3m$. A distinctive feature of this compound is the formation of a close-packed kagome bilayer unit by $Cr_3Sb$, which is sandwiched between two $Sb_2$ layers, collectively constituting a $Cr_3Sb_3$ layer (Fig. 1c). To confirm the surface termination by Sb, we examined the atomic-scale structure of the cleaved surface using scanning tunneling microscopy (STM). The STM measurements clearly reveal a triangular lattice formed by Sb atoms (Fig. 1d), with an in-plane lattice constant $a$ = 5.58 Å, in excellent agreement with X-ray diffraction results[39]. This $Cr_3Sb_3$ kagome bilayer structure sharply contrasts with other kagome materials, which usually have a uniform spacing between the kagome single layer. The unit cell of $CsCr_6Sb_6$ consists of three $Cr_3Sb_3$ kagome bilayers with ABC stacking, separated by Cs atomic layers. Notably, the distance between adjacent kagome bilayers is as large as 11.559 Å (marked as $d_1$), which is approximately four times greater than the distance between nearest Cr atoms (marked as $d_2$). The exceptionally large $d_1/d_2$ ratio in $CsCr_6Sb_6$, compared to other kagome compounds (Fig. 1e), makes it an ideal candidate for hosting flat bands with a narrow bandwidth.

To probe the flat bands, we conduct a detailed ARPES investigation of its electronic structure. Fig. 2b shows the $hv$ (photon energy)-dependent ARPES results along $\bar{\Gamma} - \bar{K}$ direction. One can see that most of the measured Fermi surfaces (FSs) do not exhibit any noticeable change with varying the incident photon energy over a wide range, indicating that these states are nondispersive along $\mathbf{k_z}$ direction. This observation directly supports the 2D nature of near-$E_F$ states in $CsCr_6Sb$, consistent with prior structural analyses.

Moving on to the FS in $\mathbf{k_x}$-$\mathbf{k_y}$ plane, Fig. 2c presents the measured constant energy contour at $E_F$ with $hv$ = 54 eV. It mainly exhibits a central spot feature at $\bar{\Gamma}$ point and six elliptical features centered around $\bar{M}$ point. Note that due to matrix element effects inherent in ARPES measurements, only half of the elliptical features are distinctly visible in Fig. 2c. Figs. 2d and 2e show the corresponding band structures along C1 and C2 directions, respectively. Both the central spot and the elliptical features at $\bar{\Gamma}$ and $\bar{M}$ points correspond to the electron pockets, labeled as $\alpha$ and $\beta$. In addition, a hole pocket emerges near $\bar{K}$ point in Fig. 2e, although its spectral weight is strongly suppressed in Fig. 2c due to photoemission matrix-element effects. While the dispersive $\alpha$, $\beta$ and $\gamma$ pockets are clearly resolved, the signature of flat band is hardly seen in Figs. 2d and 2e. To search for the flat band, we measure the overall band structure with different photon energies and polarizations. Fig. 2f presents the band structure obtained by combining LH + LV polarization, where the flat band is clearly resolved at binding energies of ~0 and -1 eV. The existence of flat bands is further confirmed by the sharp peaks in the density of states and by the low photon energy ARPES measurements (Supplementary Fig. 1). Taken together, Figs. 2d-f clearly demonstrate the coexistence of local and itinerant states near $E_F$ in CsCr$_6$Sb$_6$.

**Observation of Resonance of Flat-Band-Doublet**

Importantly, the coexistence of flat and dispersive bands allows the emergence of the long-awaited kagome flat band resonance, arising from the interaction between local and itinerant electrons. As in heavy fermion materials, a key spectroscopic feature of flat band resonance is the increase in spectral weight at low temperatures, which can be directly confirmed through temperature-dependent ARPES measurements. Figs. 3a and 3b present zoomed-in ARPES intensity plots and the corresponding second derivative plots around $\bar{\Gamma}$ point at various temperatures. At high temperatures (e.g., 92 K), only the band bottom of $\alpha$ pocket near $E_F$ is clearly visible (see Supplementary Fig. 2 for more high-temperature data). Interestingly, as the temperature decreases, three distinct flat band features (marked as f1-f3 in Fig. 3a) emerge near $\bar{\Gamma}$ point, with their characteristics most pronounced in the second derivative plot. This observation strongly suggests the occurrence of pronounced flat band resonance at low temperatures.

To further visualize the flat band resonance, we plot the energy distribution curves (EDCs) in Fig. 3c. The EDCs at 12 K exhibit a sharp coherent peak slightly below $E_F$, a key feature of flat band resonance (indicated by a blue arrow in Fig. 3c). Additionally, we resolve a satellite peak near -0.056 eV and a hump around -0.1 eV, marked by orange and black arrows, respectively. These peaks and the hump in the EDCs correspond to the three flat-band-like features observed in the ARPES intensity plot in Fig. 3a. Fig. 4a summarizes the temperature-dependent EDCs at $\bar{\Gamma}$ point. The intensities of both f1 and f2 decrease rapidly as the temperature increases near $T_N$. At 92 K, the EDCs exhibit a single broad peak near $\bar{\Gamma}$ point, indicating the disappearance of flat band resonance. This behavior cannot be explained by thermal broadening effect, as discussed in Supplementary Fig. 4. To further ensure that these changes are not caused by sample surface aging during the heating process, we performed a thermal cycle experiment. As shown in the Supplementary Fig. 5, after cooling the sample back to 12 K, the ARPES intensity plot and the second derivative plot once again exhibit the characteristic three flat-band-like features. This recovery confirms that the observed temperature-dependent evolution of the electronic structure is intrinsic to the material, rather than being an artifact of surface aging.

To quantify the resonance, we track the temperature evolution of the spectral weight associated with f1 (Figs. 4a and b). As in heavy-fermion materials, the peak evolves from incoherent at high temperature to coherent at low temperature. However, in contrast to the gradual incoherence–coherence crossover characteristic of classical $f$-electron systems, the spectral weight in $CsCr_6Sb_6$ exhibits a sharp suppression just above the short-range AFM transition. This abrupt change underscores a direct coupling between flat band resonance and magnetism, as schematically illustrated in Fig. 4b.

Notably, the flat band resonance is not confined to $\bar{\Gamma}$ point but is also observed at $\bar{M}$ point, where the electron pocket intersects the flat band (Fig. 2). Low-temperature ($T = 13$ K) ARPES measurements along the $\bar{\Gamma}$–$\bar{M}$ direction, performed using two photon energies ($h\nu = 31$ eV and $h\nu = 54$ eV), reveal clear flat band resonance features (Supplementary Fig. 7). In contrast, above $T_N$ ($T = 80$ K), the resonance vanishes, leaving only a weak-intensity flat band signal (Supplementary Fig. 7).

To gain theoretical insight, we performed both DFT (Fig. 4c and Supplementary Material S8) and DFT + DMFT calculations (Supplementary Material S9). Both approaches reproduce the flat bands near $E_F$, with orbital projections identifying dominant Cr 3$d$ contributions (Fig. 4c and Supplementary Fig. 8), thereby confirming their kagome origin. Compared with DFT, the inclusion of dynamical correlations in DMFT renormalizes the dispersive bands and further enhances their flatness[42] (Supplementary Fig. 9). In addition, the DMFT calculation captures the coherence–incoherence crossover characteristic of Kondo-like behavior[42], and the calculated local spin susceptibility highlights the involvement of local-moment screening[40], leading to a suppression of the magnetic susceptibility (upper panel in Fig. 4d). Despite these successes, notable discrepancies remain. The calculated band structure predicts additional dispersive states near $E_F$ that are not clearly resolved in ARPES, including a hole pocket below the flat band at Γ and the Dirac point at K. Surface states are excluded by the DFT calculations, which shows the overall consistency with the calculations of bulk states (Supplementary Fig.

10). Furthermore, current theoretical calculations do not account for the close interplay between flat band resonance and the short-range AFM transition observed experimentally. Bridging these gaps will require refined theoretical treatments that incorporate the influence of magnetism.

**Discussion**

The kagome bilayer structure of $CsCr_6Sb_6$ provides a unique setting for realizing the long-sought flat band resonance. First of all, as we know, flat band resonance has long eluded detection in single-kagome-layer systems, such as the well-studied 135-type compounds. This limitation arises because the single-layer configuration obstructs the essential band crossover. Specifically, in the absence of spin-orbit coupling, the kagome dispersive and flat bands remain degenerate at $\Gamma$, preventing their intersection. By contrast, $CsCr_6Sb_6$ hosts a kagome bilayer structure that generates doublet bands (Fig. 4d), allowing a flat band from one layer to intersect a dispersive band from the adjacent layer near $E_F$—setting the stage for flat band resonance. Second, the ABC stacking in $CsCr_6Sb_6$ introduces a lateral shift between bilayers, dramatically reducing interlayer orbital overlap[43] (e.g., Sb $p_z$ orbitals). Combined with a large interlayer spacing (Fig. 1e), this architecture amplifies two-dimensionality and ensures that Cr $3d$ orbitals from kagome planes dominate the low-energy electronic structure (Fig. 4c). In short, the bilayer and ABC stacking nature of $CsCr_6Sb_6$ offer an ideal playground for realizing the long-sought flat band resonance.

The kagome lattice also governs the magnetic properties of $CsCr_6Sb_6$. Magnetic susceptibility measurements show a broad anomaly around the Néel temperature ($T_N \sim 70$ K), while a Curie-Weiss analysis indicates a much larger Weiss constant ($\Theta > 600$ K)[39]. This huge disparity points to the presence of short-range AFM, which is consistent with recent muon spin relaxation measurements[40]. The kagome lattice promotes these short-range correlations through two possible mechanisms. First, theoretical studies predict that when the $E_F$ lies between the kagome flat and dispersive bands—originating from the unique kagome lattice—AFM fluctuations are substantially enhanced[6], which in turn suppresses long-range magnetic order. Second, the inherent geometric frustration of the kagome lattice, corroborated by DFT calculations showing nearly degenerate competing magnetic ground states[40], naturally favors short-range over long-range AFM order and enhances local moment fluctuations[44,45].

A defining feature of $CsCr_6Sb_6$ is the intricate interplay between flat band resonance and short-range AFM order (Fig. 4d). In conventional heavy fermion systems, suppression of long-range magnetism often paves the way for Kondo singlet formation and the development of a Kondo resonance[2,3]. $CsCr_6Sb_6$ partially echoes this paradigm: the lack of long-range AFM allows Kondo-like hybridization to coexist with the magnetism at low temperatures. Yet, unlike canonical heavy fermion metals where the Kondo resonance persists well above $T_N$ and cannot be destroyed by long-wavelength AFM fluctuations[2,3,46], in $CsCr_6Sb_6$ the resonance emerges in concert with the short-range AFM around 72 K. We attribute this synchronization to the fragility of Kondo singlets under strong local moment fluctuations[2,47,48]. In kagome systems, such fluctuations could be amplified by the flat–dispersive band interplay[6] or the frustrated magnetism[44,45], as mentioned above. Consequently, the development of short-range AFM fluctuations near $T_N$ could destabilize Kondo screening, leading to a cooperative rather than independent emergence of flat band resonance and AFM (see the lower panel in Fig. 4d). While this framework provides a plausible picture, the precise microscopic mechanism underpinning this interplay warrants further experimental and theoretical exploration.

Finally, we consider alternative mechanisms that could, in principle, account for the observed flat band resonance. Although polaron formation via electron–phonon coupling can give rise to multiple satellite features and flat-band-like spectral signatures[49], several key observations argue strongly against a polaronic origin in $CsCr_6Sb_6$ (see Supplementary Section S11). In particular, the energy separations between the resonance peaks significantly exceed the phonon cutoff frequency (Supplementary Fig. 11), and the resonance is abruptly suppressed upon warming above the short-range AFM transition. This behavior contrasts sharply with the smooth and gradual temperature evolution characteristic of polaronic effects[50]. These findings instead indicate an intrinsic connection between the flat band resonance and magnetism in the kagome lattice, allowing the polaron scenario to be excluded. Beyond electron–phonon coupling, one may also consider whether interactions between electrons and magnons could account for the observed nontrivial behavior. However, to the best of our knowledge, theoretical and experimental studies addressing such a mechanism in this context remain limited. As summarized in Fig. 4b, the logarithmic divergence of the resistivity, together with the spectroscopic observation of flat band resonance below $T_\text{N}$, provides compelling evidence that itinerant electrons couple to local magnetic moments in the ground state via Kondo interactions. Under this framework, it is unnecessary to invoke coupling to collective excitation modes such as magnons. Moreover, the existence of well-defined magnons—typically associated with long-range magnetic order—is questionable in a system exhibiting only short-range AFM correlations[51], and remains to be clarified by future studies. Nevertheless, we note that the role of magnetic excitations in correlated kagome systems remains an open issue and warrants further investigation.

Looking forward, $CsCr_6Sb_6$ presents an exciting platform for studying the interplay between flat and dispersive bands, as well as the resulting exotic quantum phases. First, as proposed in cuprates, iron-based, and heavy fermion superconductors, enhanced AFM-type spin fluctuations could mediate sign-changing Cooper pairing between different Fermi surface pockets, potentially leading to unconventional superconductivity[52]. Second, nontrivial topological phases may emerge from the spin-orbit coupling gap between the kagome flat and dispersive bands, such as Chern insulators or fractional Chern insulators[8]. Inspired by studies on magic angle twisted bilayer graphene, the quantum metric of flat bands could set a lower bound on superfluid weight, supporting the possibility of superconductivity[53,54]. Last but not least, the layered nature of $CsCr_6Sb_6$ offers an ideal platform to tune these topological and correlated phases using various techniques, including gating, doping, and strain. Notably, a recent study by Song et al.[39] has shown that the magnetism of $CsCr_6Sb_6$ exhibits remarkable layer dependence, likely due to changes in dimensionality or chemical potential. The tunability provided by the intriguing interplay between flat and dispersive bands holds great promise for the future realization of unconventional superconductivity and correlated topological phases.

In summary, we have conducted a comprehensive study of the electronic structure of the recently discovered bilayer kagome metal $CsCr_6Sb_6$. We observed the coexistence of flat and dispersive bands near $E_\text{F}$, forming the foundation for the emergence of rich and novel quantum phases in $CsCr_6Sb_6$. At low temperatures, we uncovered a strong hybridization between the flat and dispersive bands arising from doublet kagome bands of the bilayer unit, accompanied by a pronounced enhancement of spectral weight, providing the first compelling spectroscopic evidence of the long-sought flat band resonance in kagome materials. Uniquely, this resonance coincides with the previously reported onset of short-range AFM around 72 K, marking a deviation from typical Kondo lattice materials, where Kondo resonance typically occurs far above the AFM ordering temperature. Our work highlights the pivotal role of the coexistence of flat and dispersive bands in driving both kagome flat band resonance and AFM fluctuations. Moreover, it opens new avenues for engineering unconventional and nontrivial topological phases in kagome flat band systems.

## Acknowledgement

B.L. acknowledges support from the Ministry of Science and Technology of China (No. 2023YFA1407400), the National Natural Science Foundation of China (No.12374063, 92565305), the Shanghai Natural Science Fund for Original Exploration Program (No. 23ZR1479900), and the Cultivation Project of Shanghai Research Center for Quantum Sciences (No. LZPY2024). H.D. acknowledges support from the National Natural Science Foundation of China (No.12488101), the Quantum Science and Technology-National Science and Technology Major Project (No. 2021ZD0302700), the New Cornerstone Science Foundation (No. 23H010801236), and the Fundamental and Interdisciplinary Disciplines Breakthrough Plan of the Ministry of Education of China (No. JYB2025XDXM411). Y.G. acknowledges support by the National Key R&D Program of China (No. 2024YFA1400066). Y.H. acknowledges support by the Shanghai Committee of Science and Technology (No. 23JC1403300), the Shanghai Municipal Science and Technology Major Project. Z.W. is supported by the U.S. Department of Energy, Basic Energy Sciences Grant DE-FG02-99ER45747. H.T. is supported by the National Natural Science Foundation of China (No.12574270) and Science and Technology Commission of Shanghai Municipality (Pujiang Program No.24PJA051). B.J. acknowledges support by the Postdoctoral Fellowship Program of CPSF (No. GZB20250700 and 2025M773355). We thank the Shanghai Synchrotron

1. Anderson, P. W. Local Moments and Localized States. *Science* **201**, 307–316 (1978).
2. Gegenwart, P., Si, Q. & Steglich, F. Quantum criticality in heavy-fermion metals. *Nature Phys* **4**, 186–197 (2008).
3. Kirchner, S. *et al.* Colloquium: Heavy-electron quantum criticality and single-particle spectroscopy. *Rev. Mod. Phys.* **92**, 011002 (2020).
4. Checkelsky, J. G., Bernevig, B. A., Coleman, P., Si, Q. & Paschen, S. Flat bands, strange metals and the Kondo effect. *Nat Rev Mater* **9**, 509–526 (2024).
5. Chen, L. *et al.* Metallic quantum criticality enabled by flat bands in a kagome lattice. Preprint at https://doi.org/10.48550/arXiv.2307.09431 (2023).
6. Wu, S. *et al.* Flat-band enhanced antiferromagnetic fluctuations and superconductivity in pressurized $CsCr_3Sb_5$. *Nat Commun* **16**, 1375 (2025).
7. Sun, K., Gu, Z., Katsura, H. & Das Sarma, S. Nearly Flatbands with Nontrivial Topology. *Phys. Rev. Lett.* **106**, 236803 (2011).
8. Tang, E., Mei, J.-W. & Wen, X.-G. High-Temperature Fractional Quantum Hall States. *Phys. Rev. Lett.* **106**, 236802 (2011).
9. Stewart, G. R. Heavy-fermion systems. *Rev. Mod. Phys.* **56**, 755–787 (1984).

10. Zhao, W. *et al.* Gate-tunable heavy fermions in a moiré Kondo lattice. *Nature* **616**, 61–65 (2023).

11. Vaňo, V. *et al.* Artificial heavy fermions in a van der Waals heterostructure. *Nature* **599**, 582–586 (2021).

12. Wan, W. *et al.* Evidence for ground state coherence in a two-dimensional Kondo lattice. *Nat Commun* **14**, 7005 (2023).

13. Regnault, N. *et al.* Catalogue of flat-band stoichiometric materials. *Nature* **603**, 824–828 (2022).

14. Călugăru, D. *et al.* General construction and topological classification of crystalline flat bands. *Nat. Phys.* **18**, 185–189 (2022).

15. Yin, J.-X., Lian, B. & Hasan, M. Z. Topological kagome magnets and superconductors. *Nature* **612**, 647–657 (2022).

16. Wang, Y., Wu, H., McCandless, G. T., Chan, J. Y. & Ali, M. N. Quantum states and intertwining phases in kagome materials. *Nat Rev Phys* **5**, 635–658 (2023).

17. Mielke, A. Ferromagnetic ground states for the Hubbard model on line graphs. *J. Phys. A: Math. Gen.* **24**, L73 (1991).

18. Ma, D.-S. *et al.* Spin-Orbit-Induced Topological Flat Bands in Line and Split Graphs of Bipartite Lattices. *Phys. Rev. Lett.* **125**, 266403 (2020).

19. Lieb, E. H. Two theorems on the Hubbard model. *Phys. Rev. Lett.* **62**, 1201–1204 (1989).

20. Sutherland, B. Localization of electronic wave functions due to local topology. *Phys. Rev. B* **34**, 5208–5211 (1986).

21. Vergniory, M. G. *et al.* A complete catalogue of high-quality topological materials. *Nature* **566**, 480–485 (2019).

22. Neves, P. M. *et al.* Crystal net catalog of model flat band materials. *npj Comput Mater* **10**, 1–9 (2024).

23. Ye, L. *et al.* Hopping frustration-induced flat band and strange metallicity in a kagome metal. *Nat. Phys.* **20**, 610–614 (2024).

24. Huang, J. *et al.* Non-Fermi liquid behaviour in a correlated flat-band pyrochlore lattice. *Nat. Phys.* **20**, 603–609 (2024).

25. Ekahana, S. A. *et al.* Anomalous electrons in a metallic kagome ferromagnet. *Nature* **627**,

67–72 (2024).

26. Ohgushi, K., Murakami, S. & Nagaosa, N. Spin anisotropy and quantum Hall effect in the *kagomé* lattice: Chiral spin state based on a ferromagnet. *Phys. Rev. B* **62**, R6065–R6068 (2000).

27. Kang, M. *et al.* Dirac fermions and flat bands in the ideal kagome metal FeSn. *Nat. Mater.* **19**, 163–169 (2020).

28. Yin, J.-X. *et al.* Negative flat band magnetism in a spin–orbit-coupled correlated kagome magnet. *Nat. Phys.* **15**, 443–448 (2019).

29. Liu, Z. *et al.* Orbital-selective Dirac fermions and extremely flat bands in frustrated kagome-lattice metal CoSn. *Nat Commun* **11**, 4002 (2020).

30. Kang, M. *et al.* Topological flat bands in frustrated kagome lattice CoSn. *Nat Commun* **11**, 4004 (2020).

31. Di Sante, D. *et al.* Flat band separation and robust spin Berry curvature in bilayer kagome metals. *Nat. Phys.* **19**, 1135–1142 (2023).

32. Samanta, S. *et al.* Emergence of flat bands and ferromagnetic fluctuations via orbital-selective electron correlations in Mn-based kagome metal. *Nat Commun* **15**, 5376 (2024).

33. Lin, Z. *et al.* Flatbands and Emergent Ferromagnetic Ordering in $Fe_3Sn_2$ Kagome Lattices. *Phys. Rev. Lett.* **121**, 096401 (2018).

34. Ren, Z. *et al.* Persistent flat band splitting and strong selective band renormalization in a kagome magnet thin film. *Nat Commun* **15**, 9376 (2024).

35. Sun, Z. *et al.* Observation of Topological Flat Bands in the Kagome Semiconductor $Nb_3Cl_8$. *Nano Lett.* **22**, 4596–4602 (2022).

36. Gao, S. *et al.* Discovery of a Single-Band Mott Insulator in a van der Waals Flat-Band Compound. *Phys. Rev. X* **13**, 041049 (2023).

37. Yang, J. *et al.* Observation of flat band, Dirac nodal lines and topological surface states in Kagome superconductor $CsTi_3Bi_5$. *Nat Commun* **14**, 4089 (2023).

38. Jiang, Z. *et al.* Flat bands, non-trivial band topology and rotation symmetry breaking in layered kagome-lattice $RbTi_3Bi_5$. *Nat Commun* **14**, 4892 (2023).

39. Song, B. *et al.* Realization of Kagome Kondo lattice. *Nat Commun* **16**, 5643 (2025).

40. Liu, X. *et al.* Emergent dynamical Kondo coherence and competing magnetic order in a

correlated kagome flat-band metal $CsCr_6Sb_6$. Preprint at https://doi.org/10.48550/arXiv.2508.08580 (2025).

41. Shi, M. *et al.* A new class of bilayer kagome lattice compounds with Dirac nodal lines and pressure-induced superconductivity. *Nat Commun* **13**, 1–7 (2022).

42. Chen, X. & Wang, Y. Reinforcement of flat bands in a bilayer kagome metal. *Phys. Rev. B* **111**, 205120 (2025).

43. Ritschel, T. *et al.* Orbital textures and charge density waves in transition metal dichalcogenides. *Nature Phys* **11**, 328–331 (2015).

44. Burdin, S., Grempel, D. R. & Georges, A. Heavy-fermion and spin-liquid behavior in a Kondo lattice with magnetic frustration. *Phys. Rev. B* **66**, 045111 (2002).

45. Senthil, T., Vojta, M. & Sachdev, S. Weak magnetism and non-Fermi liquids near heavy-fermion critical points. *Phys. Rev. B* **69**, 035111 (2004).

46. Poelchen, G. *et al.* Unexpected differences between surface and bulk spectroscopic and implied Kondo properties of heavy fermion $CeRh_2Si_2$. *npj Quantum Mater.* **5**, 1–7 (2020).

47. Si, Q., Rabello, S., Ingersent, K. & Smith, J. L. Locally critical quantum phase transitions in strongly correlated metals. *Nature* **413**, 804–808 (2001).

48. Si, Q. Local fluctuations in quantum critical metals. *Phys. Rev. B* **68**, (2003).

49. Franchini, C., Reticcioli, M., Setvin, M. & Diebold, U. Polarons in materials. *Nat Rev Mater* **6**, 560–586 (2021).

50. Moser, S. *et al.* Tunable Polaronic Conduction in Anatase $TiO_2$. *Phys. Rev. Lett.* **110**, 196403 (2013).

51. Zhitomirsky, M. E. & Chernyshev, A. L. *Colloquium* : Spontaneous magnon decays. *Rev. Mod. Phys.* **85**, 219–242 (2013).

52. Scalapino, D. J. A common thread: The pairing interaction for unconventional superconductors. *Rev. Mod. Phys.* **84**, 1383–1417 (2012).

53. Xie, F., Song, Z., Lian, B. & Bernevig, B. A. Topology-Bounded Superfluid Weight in Twisted Bilayer Graphene. *Phys. Rev. Lett.* **124**, 167002 (2020).

54. Peotta, S. & Törmä, P. Superfluidity in topologically nontrivial flat bands. *Nat Commun* **6**, 8944 (2015).

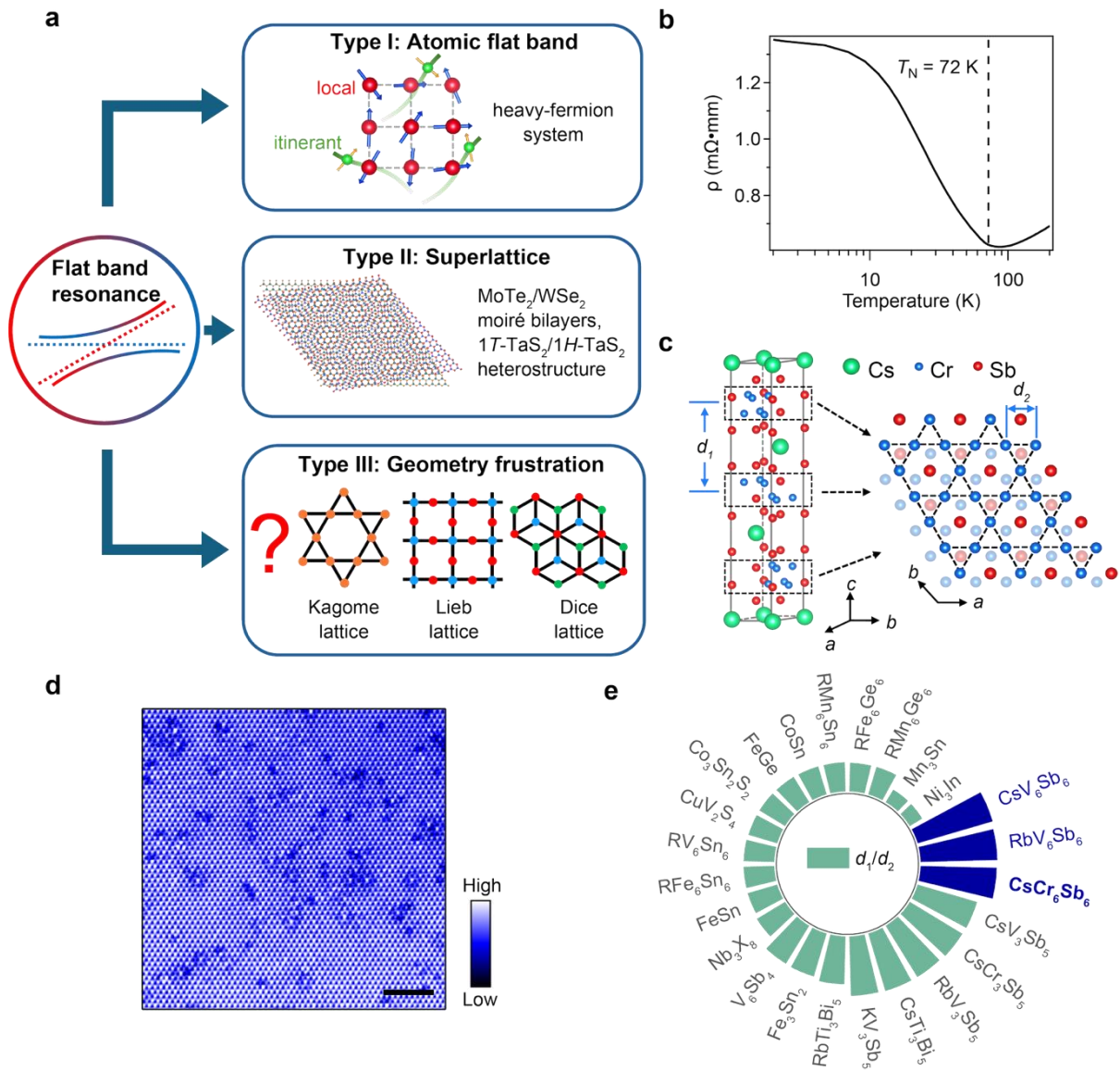

**Fig. 1 | Three types of flat band resonance and basic properties of CsCr$_6$Sb$_6$. a**, A schematic illustration of three types of flat band resonances. In Type I materials, the flat band resonance originates from the interactions between local electrons (red spheres), which is typically 4*f*/5*f* electrons in heavy fermion systems[9], and conduction electrons (green spheres). Blue and yellow arrows indicate the magnetic moment, while the green traces represent the itinerancy of the conduction electrons. Type II materials consist of van der Waals superlattice，such as MoTe$_2$/WSe$_2$ moiré bilayers[10] and 1T-TaS$_2$/1H-TaS$_2$ heterostructure[11]. In type III materials, flat bands emerge due to quantum geometry frustration in specific lattices，such as kagome[15–17], Lieb[19] and dice lattices[20]. However, the flat band resonance in type III materials has not yet been experimentally discovered. **b**, Resistivity of CsCr$_6$Sb$_6$, presenting a weak kink at 72 K consistent with previously reported onset of short-range antiferromagnetic (AFM) order. **c,**

Crystal structure of $CsCr_6Sb_6$. Dashed rectangle indicates the kagome bilayer, which is emphasized in the right panel. $d_1$ and $d_2$ indicate the distance between adjacent bilayer and nearest Cr atoms, respectively. **d,** STM topographic image obtained from the Sb surface [$T$ = 4.2 K, (V, I) = (20 mV, 100 pA), image size: 30 nm × 30 nm, scale bar: 5 nm]. **e,** the $d_1/d_2$ ratio of various of kagome materials experimentally reported, see details in Supplementary Table 1. The large $d_1/d_2$ ratio of $CsCr_6Sb_6$, compared with other materials, demonstrates the ideal two dimensionality.

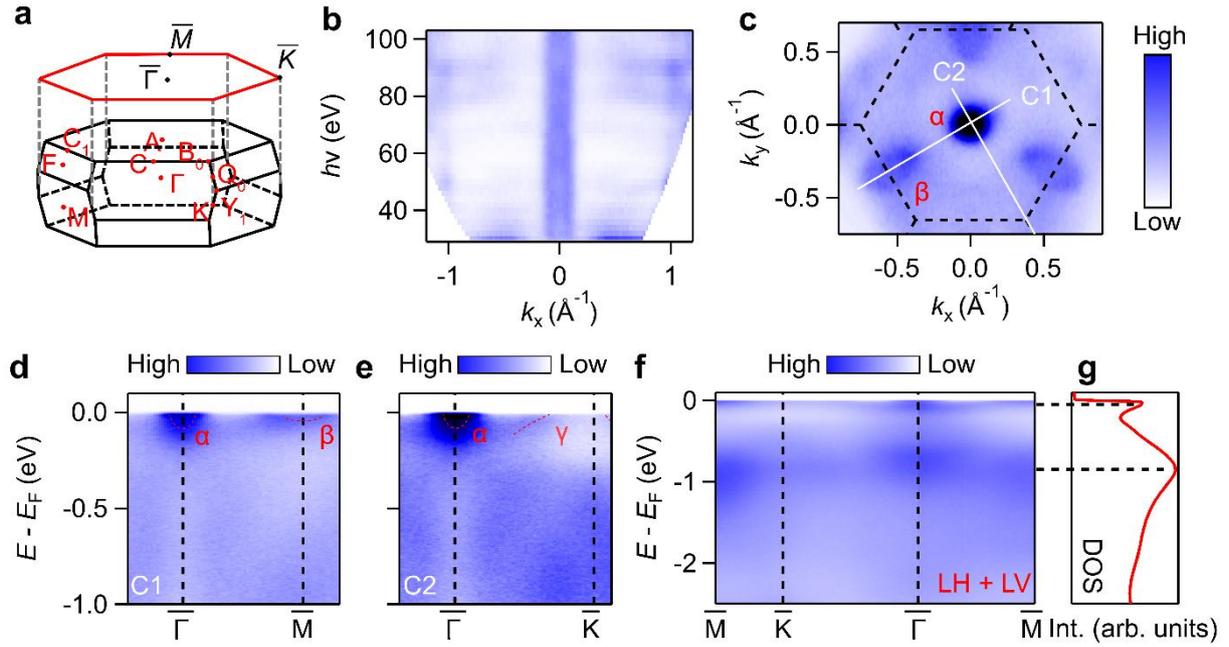

**Fig. 2 | Coexistence of flat bands and dispersive bands at $E_F$ in CsCr$_6$Sb$_6$. a,** Bulk and projected surface Brillouin zones. **b,** ARPES intensity plot of constant energy contour at $E_F$ measured along $\bar{\Gamma}$-$\bar{K}$ at 15 K with varying photon energies. **c,** ARPES intensity plot of constant energy contour at $E_F$, showing the FS in the $\mathbf{k_x}$-$\mathbf{k_y}$ plane. **d, e,** ARPES intensity plot along the high symmetry lines C1 and C2, as indicated white lines in c, respectively. **f,** ARPES intensity plots along $\bar{M} - \bar{\Gamma} - \bar{K} - \bar{M}$. The spectra are obtained by summing the data collected under linear horizontal (LH) and linear vertical (LV) polarizations. **g,** Density of states extracted

from the ARPES data in **f**. Data in **b–e** were collected with LH polarization, and data in **c-g** are collected at 15 K with $h\nu$ = 54 eV.

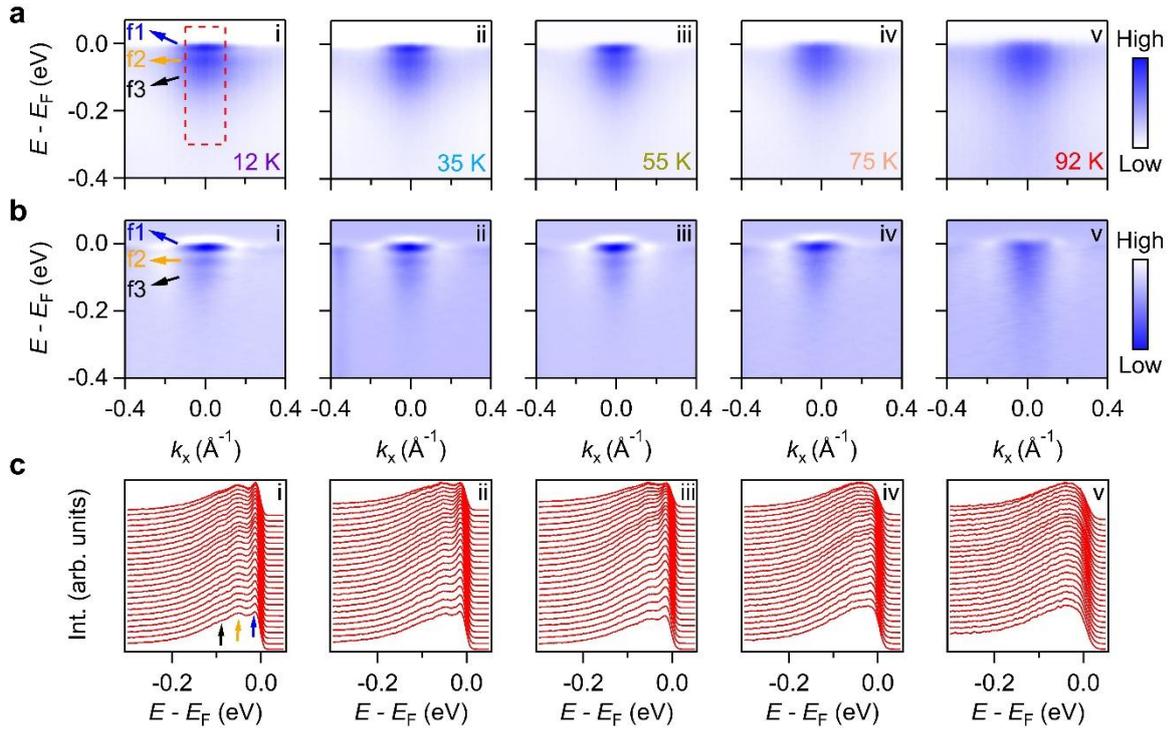

**Fig. 3 | Kagome flat band resonance in CsCr$_6$Sb$_6$. a**, ARPES intensity plots around $\bar{\Gamma}$ point along $\bar{\Gamma} - \bar{K}$ measured with $h\nu$ = 54 eV under LH polarization at various temperatures. f1, f2 and f3 indicate three flat band features. **b**, The second derivative of the ARPES intensity plots shown in **a**, obtained by combining derivatives along both energy and momentum directions. The uncombined second derivative plots are also presented in Supplementary Fig. 3. **c**, EDCs extracted from the ARPES data within the red rectangle in **a**. Blue, orange and black arrows mark the coherent peak, satellite peak, and hump in EDCs, respectively.

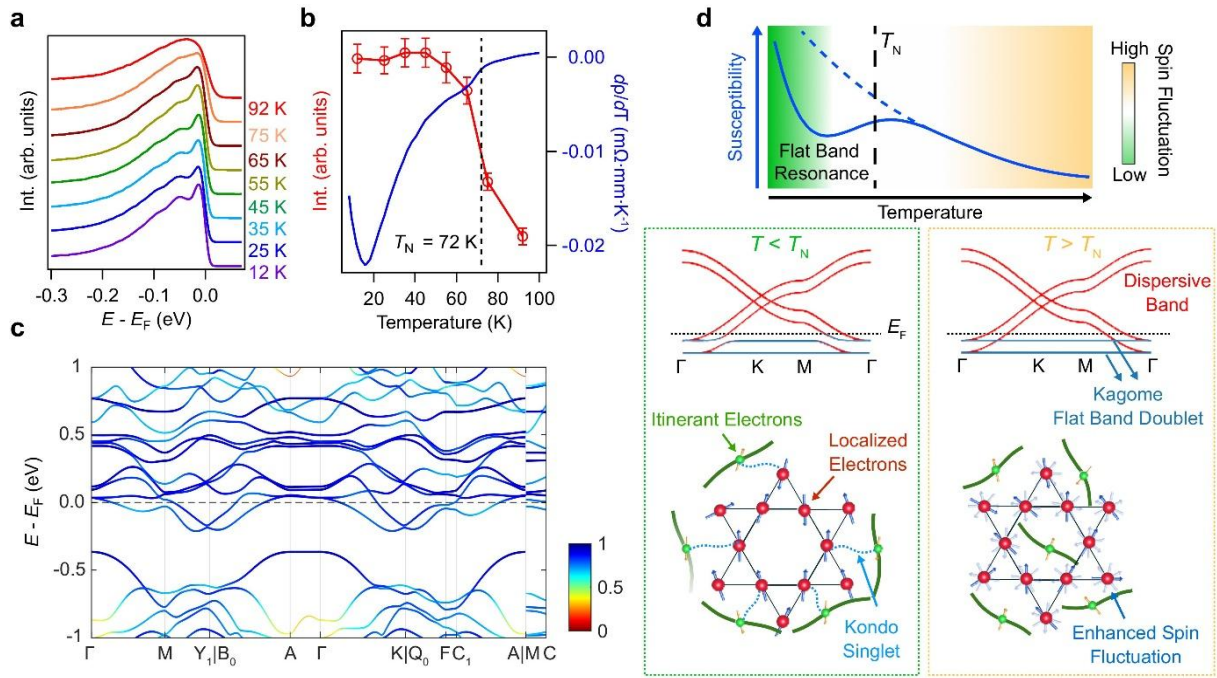

**Fig. 4 | Evolution of the kagome flat band resonance. a,** Comparison of EDCs at $\bar{\Gamma}$ point at different temperatures, showing the disappearance of coherent peak above 75 K. **b,** Temperature dependence of the coherent peak spectral weight (red curve), revealing a phase transition around 72 K. Error bars represent ten times the standard deviation of the fitting process, as detailed in Supplementary Fig. 6. The derivative of resistivity with respect to temperature (blue curve) exhibits a kink at approximately 72 K, indicating the onset of short-range AFM order. **c,** Orbital resolved band structure of $CsCr_6Sb_6$ along high symmetry lines based on DFT calculation, where the color scale indicates the contribution of Cr 3$d$ orbitals. **d,** Schematic illustration of the kagome flat band resonance in $CsCr_6Sb_6$. The kagome bilayer units in the crystal structure give rise to flat band doublets and dispersive bands near $E_F$. The interplay between them enhances the spin fluctuation, resulting in the emergence of short-range AFM order at $T_N$. Upper panel schematically illustrates that the onset of short-range AFM order leads magnetic susceptibility (blue solid line, taken from the experimental results) to deviate from the Curie-Weiss behavior (blue dashed line, extrapolated from the high temperature side) at low temperatures. Below $T_N$, short-range AFM order stabilizes local magnetic moments, facilitating Kondo singlet formation, which manifests as a kagome flat band resonance in the spectral weight. Above $T_N$, strong AFM fluctuations hinder the formation of stable local magnetic moments, preventing the emergence of Kondo singlets and suppressing hybridization between flat and dispersive bands.

## Methods

**Single crystal growth and basic characterizations.** Single crystals were grown by using the self-flux method. Starting materials of high-purity Cs liquid (99.98%), Cr powder (99.9%) and Sb granules (99.999%) were mixed in a molar ratio of 1: 1: 5, placed into an alumina crucible and was then sealed into a tantalic tube under vacuum. The sealed tube was heated to 1100°C within 15 hours and subsequently cooled down to 900°C within 5 hours, kept at that temperature for 10 hours and slowly cooled to 500°C at a rate of 1.5°C/h. Then the tube was cooled to room temperature in air, the excess flux is removed using ultrapure water to etch away the Cs-Sb phases. Single crystal X-ray diffraction (SXRD) was performed using a Bruker D8 single crystal X-ray diffractometer with Mo Kα1 source ($\lambda$ = 0.71073 Å) at 298 K. The composition of the crystals was examined using energy-dispersive x-ray spectroscopy (EDS). Electrical transport measurements were conducted using a Quantum Design Physical Property Measurement System. A standard four-probe method was employed to measure resistivity, with the electric current applied within the *ab*-plane.

**Angle-resolved photoemission spectroscopy**. High resolution ARPES measurements were performed at the "Dreamline" (BL09U) beamline and the BL03U beamline of the Shanghai Synchrotron Radiation Facility (SSRF), using a Scienta Omicron DA30L electron analyzer. The data were collected over a photon energy range of 30 eV to 102 eV, and the combined (beamline and analyzer) experimental energy resolution and was set to 15 meV. The angular resolution of the DA30L analyzer was 0.1°. The samples were cleaved *in situ* under a base pressure better than $8 \times 10^{-11}$ mbar.

**DFT calculation.** Electronic structure calculations were performed using density functional theory (DFT) as implemented in the Vienna Ab-initio Simulation Package (VASP). Projector-augmented wave (PAW) potentials are employed, considering 9 valence electrons for Cs, 6 for Cr, and 5 for Sb. The generalized gradient approximation (GGA) parameterized by Perdew-Burke-Ernzerhof (PBE) was used to describe the exchange-correlation interaction. A plane-wave energy cutoff of 300 eV was applied, and the Brillouin zone is sampled with a 6×6×6 k-point mesh. All calculations were based on the experimental crystal structure of $CsCr_6Sb_6$.

**DFT+DMFT calculation.** We studied the correlation effects via charge self-consistent

DFT+DMFT using Wien2k (GGA-PBE, $RK_{max}$ = 7.0, 10×10×10 k-mesh) and the embedded-DMFT package. The impurity problem for Cr-3d orbitals (U = 3.5 eV, J = U/5) was solved by continuous-time quantum Monte Carlo (CT-QMC), with spectral functions computed via maximum entropy analytic continuation.

**STM measurement**. The STM experiments were performed using a commercial Unisoku USM1300 low-temperature STM machine. Pt/Ir tips were used, and conditioned by field emission with a gold target. To obtain clean surface for STM measurements, the samples were cleaved in-situ at ~ 15 K in ultrahigh vacuum (base pressure ≈ $2 \times 10^{-10}$ mbar), then transferred immediately to the STM stage (maintained at 4.2 K) for STM measurements.

## Data Availability

The data generated in this study have been deposited in the Zenodo database at https://doi.org/10.5281/zenodo.18561671.